\documentclass[twoside,fleqn]{article}
\usepackage{espcrc2}

\title{Corrections to finite-size scaling in two-dimensional
$O(N)$ $\sigma$-models
}

\author{
  Sergio Caracciolo\address{Dipartimento di Fisica and INFN,
      Universit\`a degli Studi di Lecce,
      Lecce 73100, ITALIA}
 $\!$and
  Andrea Pelissetto\address{Dipartimento di Fisica and INFN,
      Universit\`a degli Studi di Pisa,
      Pisa 56100, ITALIA}
}

\begin{document}

\begin{abstract}
We have considered the corrections to the finite-size-scaling functions
for a general class of $O(N)$ $\sigma$-models with two-spin interactions
in two dimensions for $N=\infty$. We have computed the leading corrections
finding that they generically behave as $(f(z) \log L + g(z))/L^2$ where
$z = m(L) L$ and $m(L)$ is a mass scale; $f(z)$ vanishes for Symanzik improved 
actions for which the inverse propagator behaves as $q^2 + O(q^6)$ for 
small $q$, but not for on-shell improved ones. We also discuss a model
with four-spin interactions which shows a much more complicated 
behaviour. 
\end{abstract}

\maketitle

\newcommand{\reff}[1]{(\ref{#1})}
\def\smfrac#1#2{{\textstyle\frac{#1}{#2}}}

\newcommand{\be}{\begin{equation}}
\newcommand{\ee}{\end{equation}}
\newcommand{\<}{\langle}
\renewcommand{\>}{\rangle}

\def\bsigma{\mbox{\protect\boldmath $\sigma$}}
\def\btau{\mbox{\protect\boldmath $\tau$}}

We have studied the finite-size-scaling (FSS) behaviour of 
$O(N)$ models for large $N$. The purpose of the study was twofold:
first of all we wanted to understand how mean values computed on a finite 
lattice of size $L$ converge to their infinite-volume values; moreover 
we wanted to understand the functional form of the correction to FSS.
For this purpose we have considered a generic hamiltonian of the 
form 
\be 
H\, =\, N\beta \sum_{xy} J(x-y) \,\bsigma_x\cdot\bsigma_y\;\; .
\label{hamiltonian}
\ee
The coupling is required to be local so that the Fourier 
transform $\hat{J}(q)$ is continuous. Moreover, to have the correct
continuum limit we require $\hat{J}(q) - \hat{J}(0)$ to have a unique 
zero in the Brillouin zone at $q=0$ with $\hat{J}(q) - \hat{J}(0)\sim q^2$
for $q\to 0$. 

We have considered the $N\to\infty$ limit of the model defined by 
\reff{hamiltonian} at fixed $\beta$ in a finite box of size 
$L\times T$ and in a strip of width $L$, in both cases using 
periodic boundary conditions  in the finite direction(s), and we have 
studied the FSS limit $L\to\infty$, $T\to\infty$ with 
$T/L\equiv \rho$ fixed ($\rho=\infty$ for the strip) and 
$m(L,T) L$ fixed where $m(L,T)$ is (some) mass scale.

On a finite lattice the theory is parameterized by a mass $m_{L,T}$ related 
to $\beta$ by 
\be
\beta\, =\, {1\over LT} \sum_{p_x,p_y} {1\over w(p) + m^2_{L,T}} \, 
\equiv {\cal I}_{L,T}(m^2_{L,T} )
\label{gapeq}
\ee
where $w(p) = 2 (\hat{J}(q) - \hat{J}(0))/\hat{J}''(0)$ and the sum extends 
over the points $(p_x,p_y) = 2 \pi(n_x/L,n_y/T)$, $0\le n_x\le L-1$ and 
$0\le n_y\le T-1$. When the lattice is infinite in one (or both)
direction(s) the sum is substituted by the corresponding integral.

The basic tool for the computation of the FSS functions is the expansion in
the FSS limit of the r.h.s of \reff{gapeq}. We find a general result of the 
form
\begin{eqnarray}
&& \hskip -20pt 
   {\cal I}_{L,T} (m^2_{L,T} ) = 
   {1\over 2\pi} \log L +\, F_0(z;\rho) + \Lambda_0 \nonumber \\
&& + {z^2\over 4 \pi L^2} (3 \delta_1 + 4 \delta_2 ) \log L + 
   {1\over L^2} F_1(z;\rho) 
\label{expansionI}
\end{eqnarray}
where $z \equiv m_{L,T} L$ and the neglected terms are of order 
$O(\log L/L^4)$. We want to make a few general remarks on this result:
\begin{enumerate} 
\item The leading logarithm is universal and it is indeed connected to the 
leading term of the $\beta$-function.
\item $F_0(z;\rho)$ is a universal function (i.e. independent on the 
  specific $J(x)$) which depends only on the modular parameter $\rho$.
\item In the leading (for $L\to\infty$) term of the expansion, the only 
  dependence on $J(x)$ is due to $\Lambda_0$ given by 
\be
   \Lambda_0 =\, \int {d^2p\over (2 \pi)^2} \left( 
     {1\over w(p)} - {1\over\hat{p}^2}\right)  
\ee
where $\hat{p}^2 = 4 \sum_i \sin^2 (p_i/2)$.
\item The $\log L/L^2$ term is {\em not} universal but depends only on the 
  small-$q^2$ behaviour of $w(q)$; indeed it depends on $\delta_1$ and 
  $\delta_2$ defined by
\be 
  w(q) \, =\, q^2 + \delta_1 \sum_\mu q^4_\mu + \delta_2 (q^2)^2 + O(q^6)
\ee
\end{enumerate}
Given the expansion \reff{expansionI} it is easy to obtain the FSS 
functions (including the corrections of order $O(1/L^2)$) of various 
quantities. If one considers the {\em true} mass gap $\mu(L)$ on a 
strip of width $L$ (this quantity cannot obviously be defined on a 
$L\times T$ lattice) one obtains in terms of $x=\mu(L) L$ 
\begin{eqnarray}
&& \hskip -20pt 
   \left( {\mu(\infty)\over\mu(L)}\right)^2\, =\, 
{32\over x^2} e^{-4\pi F_0(x;\infty)} \times \nonumber \\
&& \hskip -20pt
   \left(1 + {\Delta_1(x)\over L^2} \log L + {\Delta_2(x)\over L^2} + 
   O(L^{-4} \log L)\right)
\end{eqnarray}
where $\Delta_1(x)$ and $\Delta_2(x)$ have been explicitly computed. The
leading contribution is in agreement with the result by L\"uscher
\cite{Luscher}.
The function $\Delta_1(x)$ is very simple: explicitly
\be
\Delta_1(x) =\, (3\delta_1 + 4\delta_2) (32 e^{-4 \pi F_0(x;\infty)} 
    - x^2) 
\ee
and thus this term vanishes for Symanzik actions for which 
$\delta_1=\delta_2=0$, but not for on-shell improved actions for which 
$\delta_1=0$ but $\delta_2$ is arbitrary. In particular it does not vanish 
for the so-called ``perfect" laplacian \cite{Wilson,Hasenfratz} 
unless $\kappa = 4$.
Notice moreover that in the perturbative (PT) regime
this term is proportional to $x^2$ and thus appears only 
in {\em two}-loop PT computations. For the function $\Delta_2(x)$ we
obtain in the PT limit $x\to 0$
\be
\Delta_2(x) =\,  {2 \pi^2\over3}\delta_1 - 4 \pi (\delta_1 + \delta_2) x 
   + O(x^2)
\label{Delta2}
\ee
Thus the tree-level correction vanishes for $\delta_1=0$ (the so-called 
{\em on-shell} improved actions) independently of $\delta_2$. The vanishing 
of the second (one-loop) term requires instead $\delta_1+\delta_2=0$: 
this is verified by Symanzik actions for which $\delta_1=\delta_2=0$ 
(and in this case the one-loop corrections behave as $1/L^4$) but 
not by generic {\em on-shell} improved actions. Notice however that 
this cancellation does {\em not} happen 
for the complete function and 
thus the corrections to FSS always behave as $1/L^2$ (it is for instance
easy to see that terms of order $x^2$ do {\em not} cancel in 
$\Delta_2(x)$ even for $\delta_1=\delta_2=0$).

Completely analogous formulae can be derived for other observables on the 
strip. In particular one can consider the inverse second-moment 
correlation length: the result is similar to that of the true mass gap $\mu$,
the only difference being the function $\Delta_2(x)$. However, 
up to $O(x^2)$ its expansion is still given by \reff{Delta2}. 

Similar results are valid on a torus. In this last case we take as variable 
$z\equiv m_{L,T} L$, $m_{L,T}$ being the mass appearing in the 
gap equation \reff{gapeq} which is the inverse of the second-moment
correlation length $\xi^{(2)}(L,T)$.

It is interesting to consider the limit $z\to\infty$ (infinite-volume
limit): in this case we find generically, as expected,
\be
{ {\cal O}(L,T)\over {\cal O}(\infty)}\, =\, 
  1 + O(z^p e^{-z}) + O(\rho^p z^p e^{-\rho z}) 
\label{behaviourlargez}
\ee
i.e. the infinite-volume limit is essentially reached exponentially
($p$ is an exponent which depend on the specific observable ${\cal O}$). 
A notable 
exception is the second-moment correlation length which has corrections 
of order $O(1/z^2)$. The reason is essentially due to the fact 
that the standard definitions of $\xi^{(2)}$ on a finite volume are such that,
for $L\to\infty$, 
\be
\xi^{(2)}(L) = \xi^{(2)}(\infty) +\, O(1/L^2) 
\ee
These $1/L^2$ terms are the cause of the $1/z^2$ corrections in the 
FSS functions. It is however possible to define a correlation length
which does not suffer from this problem: define 
$\tilde{\xi}^{(2)}(L,T)$ by (assume for simplicity $L$ and $T$ even) 
\begin{eqnarray}
&& \hskip -20pt 
   \left(\tilde{\xi}^{(2)}(L,T)\right)^2 \,=\, \nonumber \\
&& {1\over 4 \hat{G}(0)} 
   \sum_{i=1-L/2}^{L/2} \sum_{j=1-T/2}^{T/2} (i^2+j^2) G(i,j)
\end{eqnarray}
for any two-point function $G(x)$ (here $\hat{G}$ denotes the Fourier transform
of $G$). It is easy to see that $\tilde{\xi}^{(2)}(L,T)\to \xi^{(2)}(\infty)$
exponentially and that, for $z\to\infty$, 
the corrections are of order $O(e^{-\rho z/2},e^{-z/2})$.

We have also considered a more general class of models of the form
\be
H\,=\, N\beta \sum_{\<xy\>} \left[ (1-r)  \bsigma_x\cdot\bsigma_y + 
   {r\over2} \left(\bsigma_x\cdot\bsigma_y\right)^2 \right]
\label{mixed}
\ee
with nearest-neighbour interactions only (this is the only case which can 
be easily solved in the large-$N$ limit \cite{Magnoli}). 
Here $r$ is a free parameter which interpolates between the standard
$N$-vector hamiltonian ($r=0$) and the $RP^{N-1}$ standard hamiltonian
($r=1$).
We wanted indeed to understand 
if the functional form of the corrections to FSS we have found for 
generic two-spin interactions is general or not: one reason to believe 
that two-spin interactions can give rise to a simpler behaviour is the 
fact, for instance, that the $\beta$-function for models defined 
by \reff{hamiltonian} has only the leading term, all others vanishing.
Instead, for $r\not=0$, the hamiltonian \reff{mixed} has a $\beta$-function
which is non vanishing to all orders in $1/\beta$. And indeed for 
the models defined by \reff{mixed} we have found a more complicated 
behaviour. Considering for instance the mass gap in a strip
we have found
\begin{eqnarray}
&& \hskip -20pt \left({\mu(\infty)\over\mu(L)}\right)^2\, =\, 
{32\over x^2} e^{-4\pi F_0(x;\infty)} \times \nonumber \\
&& \hskip -10pt 
   \left(1 + {\Delta_1(x)\over L^2} \log L + {\Delta_2(x)\over L^2} + 
   r {\Delta_3(x,L)\over L^2}\right)
\end{eqnarray}
with corrections of order $O(L^{-4}\log L)$. Here 
\be
\Delta_3(x,L)\, =\, \overline{\Delta}_3(x) 
   {(\log L + 2\pi F_0(x;\infty))^2\over \log L + 2\pi F_0(x;\infty)-2r}
\label{delta3}
\ee
As expected the FSS function is universal ($r$-independent). 
The new feature is the appearance of the term \reff{delta3}: in the limit 
$L\to\infty$ at $z$ fixed, this new term behaves as $\log L/L^2$ and thus 
the leading correction has the same functional form of the case 
we considered before; however in this case beside $1/L^2$ corrections there 
are also $1/(L^2 \log L)$ terms and so on. The origin of these terms 
can be understood if we expand them in the PT limit, $z\to 0$ with 
$L$ {\em fixed} (notice that here we are making an {\em illicit} exchange of 
limits). Since $F_0(x;\infty) = 1/(2 x) (1 + O(x))$, 
we obtain an expansion on the form 
\begin{eqnarray}
&& \hskip -20 pt 
  {1\over L^2} [ b_{00} + x (b_{11} \log L + b_{10})  \nonumber \\
&& \hskip -8pt 
    + x^2 ( b_{22} \log^2 L + b_{21} \log L + b_{20}) + O(x^3 ) ]
\end{eqnarray}
which is indeed the pattern one expects from multiloop sums.
An $n$-loop sum will generically behave as 
\be
P_n(\log L) +\, Q_n(\log L)/L^2 + O(L^{-4} \log^n L )
\ee
where $P_n(y)$ and $Q_n(y)$ are $n$-degree polynomials. 
Of course to obtain the correct corrections to FSS one must resum 
the PT expansion in order to be able to exchange the limits 
(the PT expansion is obtained for $z\to 0$ at $L$ fixed and we want 
to obtain the behaviour for $L\to\infty$ for small, but fixed $z$). 
Our calculation shows that these infinite series of logarithms
resum giving rise to corrections which still behave as $\log L/L^2$. 
Of course, it would be interesting to know if this is still true for 
generic models ($N=\infty$ models usually have a behaviour which is 
simpler than that of generic ones): indeed one could be worried 
by the possibility that the terms in $\log L/L^2$ resum non trivially 
to give corrections with a non trivial power, i.e. corrections
to FSS behaving as $L^p/L^2$, $p>0$. Unfortunately we do not have any answer 
to this question.

A detailed presentation will appear in a forthcoming paper \cite{CP}.

%
%
%%%%%%%%%%%%   references  %%%%%%%%%%%%%%%%%%%%%%%%
%

\end{document}